\documentclass[aps,prl,amsmath,twocolumn,showpacs,superscriptaddress]{revtex4-1}  
\usepackage{graphicx}  
\usepackage{dcolumn}   
\usepackage{amssymb}   

\usepackage{color}
\usepackage[usenames,dvipsnames,svgnames]{xcolor}
\usepackage[utf8]{inputenc}
\usepackage{tabularx}
\usepackage{ragged2e}
\usepackage{amsmath}
\usepackage{bm} 
\usepackage{mathtools}
\usepackage{afterpage}

\DeclarePairedDelimiter\abs{\lvert}{\rvert}
\newcommand{\difdisp}[2]{\frac{\mathrm{d}#1}{\mathrm{d}#2}} 
\newcommand{\pdifdisp}[2]{\frac{\partial #1}{\partial #2}}

\newcommand{\imi}[0]{\mathrm{i}} 

 \DeclareFontFamily{U}{wncy}{}
    \DeclareFontShape{U}{wncy}{m}{n}{<->wncyr10}{}
    \DeclareSymbolFont{mcy}{U}{wncy}{m}{n}
    \DeclareMathSymbol{\Sh}{\mathord}{mcy}{"58} 

\begin{document}


\hspace{5.2in} \mbox{Fermilab-Pub-04/xxx-E}

\title{Self-synchronization of Kerr-nonlinear Optical Parametric Oscillators}

\author{Hossein Taheri}
\affiliation{School of Electrical and Computer Engineering, Georgia Institute of Technology, Atlanta GA, USA}
\author{Pascal Del'Haye}
\affiliation{National Physical Laboratory, Teddington, UK}
\author{Ali A. Eftekhar}
\affiliation{School of Electrical and Computer Engineering, Georgia Institute of Technology, Atlanta GA, USA}
\author{Kurt Wiesenfeld}
\email{kurt.wiesenfeld@physics.gatech.edu}
\affiliation{Center for Nonlinear Science, School of Physics, Georgia Institute of Technology, Atlanta GA, USA}
\author{Ali Adibi}
\email{ali.adibi@ece.gatech.edu}
\affiliation{School of Electrical and Computer Engineering, Georgia Institute of Technology, Atlanta GA, USA}

\date{\today}

\begin{abstract}
We introduce a new, reduced nonlinear oscillator model governing the spontaneous creation of sharp pulses in a damped, driven, cubic nonlinear Schroedinger equation.  The reduced model embodies the fundamental connection between mode synchronization and spatiotemporal pulse formation.  We identify attracting solutions corresponding to stable cavity solitons and Turing patterns.  Viewed in the optical context, our results explain the recently reported $\pi$ and $\pi/2$ steps in the phase spectrum of microresonator-based optical frequency combs.
\end{abstract}

\pacs{05.45.Xt, 05.65.+b, 42.65.Re, 42.65.Sf, 42.65.Tg}
\maketitle

\begin{figure}[htbp]
\centering
\includegraphics[width=0.49\textwidth]{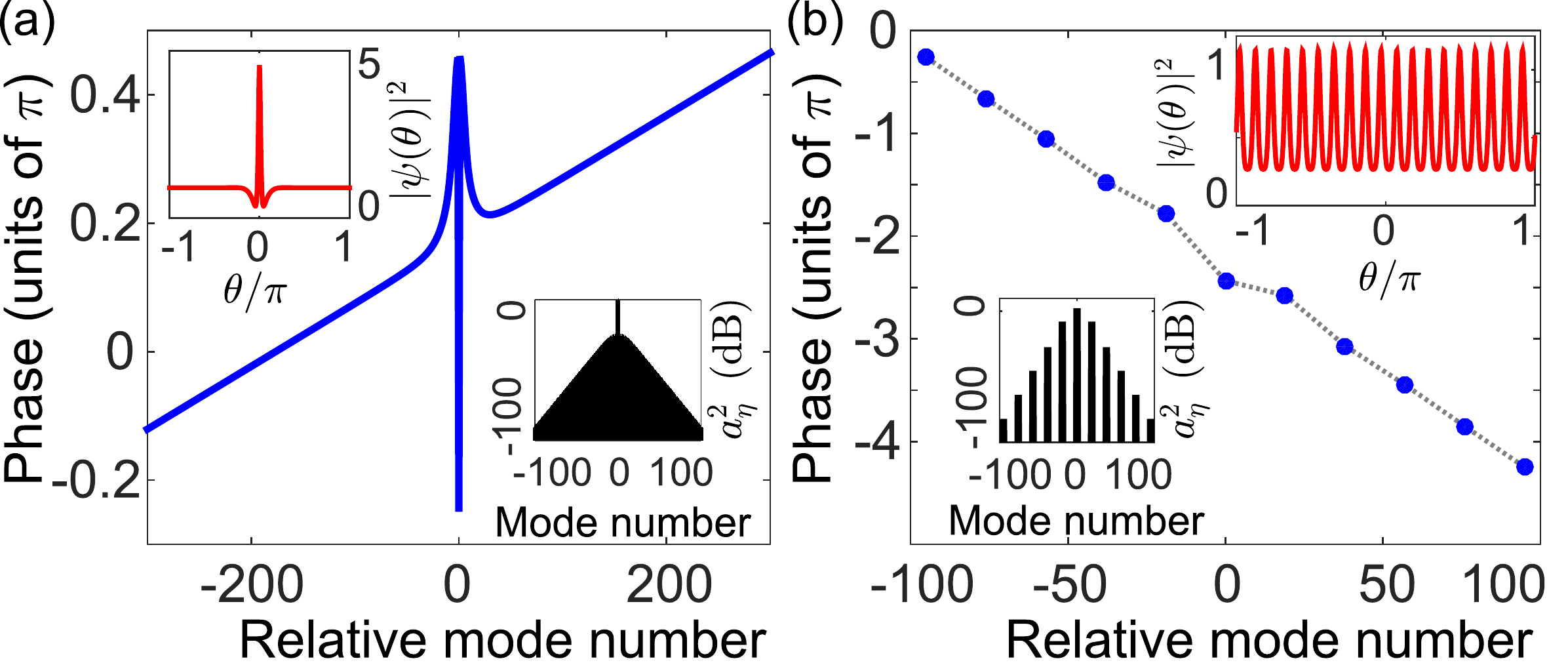}
\caption[Phase alignment in (a) solitons and (b) Turing rolls.]{\label{fig:prelim:phasealigned}Phase alignment in (a) solitons and (b) Turing rolls seen in the steady-state solutions of Eqs.~(\ref{eq:LLE}) and (\ref{eq:CNODE}). The inset curves in red (top corners) show the spatiotemporal waveforms and those in black (bottom corners) are the frequency spectra. For both solitons and rolls the phases lie on straights lines of arbitrary slope. Parameter values are (a) $\alpha=2,\, d_2=-0.0124,\, F=1.41$, and (b) $\alpha=0,\, d_2=-0.0124,\, F=1.63$. The phase profile has been unwrapped in (b).}
\end{figure}

Unlike pulsed lasers, pulsation in optical microresonators requires neither active nor passive mode locking elements (e.g. modulators or saturable absorbers)\cite{haus2000mode, kutz2006mode}. Rather, pulsed states arise naturally from a simple damped driven nonlinear Schroedinger equation known as the Lugiato-Lefever equation (LLE) \cite{lugiato1987spatial, haelterman1992dissipative, matsko2011mode, coen2013modeling, chembo2013spatiotemporal}. Two categories of stable pulsed state solutions have been identified for the LLE: stable modulational instability (MI, also known as hyper-parametric oscillations or Turing rolls) and stable cavity solitons \cite{matsko2012hard, erkintalo2014coherence}. Turing rolls arise from the intra-cavity equilibrium field through modulational instability of vacuum fluctuations and usually have multiple-FSR (free spectral range) spacing between their adjacent teeth, while the word soliton is used to refer to coherent combs with single-FSR spacing.  Experimental and theoretical studies have suggested that soliton states are not accessible from the continuous wave (CW) intra-cavity field without seeding \cite{taheri2015soliton, lobanov2015generation}, changing the pump frequency or power \cite{matsko2012hard, lamont2013route, herr2014temporal}, or a suitable input pulse \cite{leo2010temporal}. Owing to the low phase noise and exceedingly stable frequency spacing of the comb teeth in Turing rolls and solitons, chip-scale pure low-phase-noise radio frequency (RF) sources \cite{liang2015high} and coherent communication with speeds in excess of 100 Gbit/s per comb line have been demonstrated \cite{pfeifle2014coherent, pfeifle2015optimally}.\newline
\indent In addition to the generation of a frequency comb with equidistant teeth, temporal pulse generation requires mutual phase locking of the complex amplitudes. Phase locking in optical microresonators has been studied in terms of the cascaded emergence of phase-locked triplets \cite{coillet2014robustness}. Injection locking of overlapping bunched combs has been explained using the Adler equation \cite{del2014self}. Few-mode models have explained the phase offset between the pumped mode and the rest of the comb teeth \cite{loh2014phase, taheri2015anatomy}. More recently, Wen \emph{et al.} \cite{wen2014self} have emphasized the link between oscillator synchronization---most famously described by the Kuramoto model \cite{strogatz2000kuramoto}---and the onset of pulsing behavior. However, while stable ultrashort pulses have been demonstrated in a variety of microresonator platforms \cite{herr2014temporal, saha2013modelocking, brasch2016photonic, vahala2015soliton}, the underlying phase locking mechanism is still unknown. As a result, features of microcomb phase spectra revealed in recent measurements \cite{del2015phase} are yet not understood.\newline
\indent In this paper, we introduce a reduced phase model which governs the nonlinear mode interactions responsible for spontaneous creation of pulsed states in the LLE which result from a balance between Kerr nonlinearity, dispersion (or, in the spatial case, diffraction), parametric gain, and cavity loss \cite{grelu2012dissipative}.  The model interactions are \emph{ternary} (that is, they involve three-variable combinations) rather than binary, as in typical phase models.  Our model admits attracting solutions which, interpreted in the context of nonlinear optical cavities, correspond to stable cavity solitons and Turing patterns, and provides an explanation of recent observations of phase steps in optical frequency combs. Moreover, our model clarifies the role of MI and chaos in the generation and stability of Turing rolls and solitons. \newline
\indent The LLE is a nonlinear partial differential equation in time and the azimuthal angle around the whispering-gallery mode resonator \cite{chembo2013spatiotemporal}, or, equivalently, in a slow and a fast time variable \cite{haelterman1992dissipative,coen2013modeling}. Equivalently, a set of coupled nonlinear ordinary differential equations (ODEs) can be used to study resonator-based optical frequency comb generation \cite{chembo2010modal}. The generalized spatiotemporal LLE in normalized form
\begin{equation}\label{eq:LLE}
\pdifdisp{\psi}{\tau}=-(1+\imi\alpha)\psi-\imi \frac{d_2}{2} \frac{\partial^2\psi}{\partial\theta^2}+\imi|\psi|^2\psi+F,
\end{equation}
and its corresponding ODEs
\begin{equation}\label{eq:CNODE}
\difdisp{\tilde{a}_\eta}{\tau}=-(1+\imi\alpha)\tilde{a}_\eta+\imi\frac{d_2}{2}\eta^2\tilde{a}_\eta+\imi\sum_{l,\, m,\, n}\tilde{a}_l \tilde{a}_m^*\tilde{a}_n \, \delta_{\eta_{lmn}\eta}+\tilde{F}_\eta,
\end{equation}  
are Fourier transform pairs, the conjugate variables of the transform being the azimuthal angle around the resonator $\theta$ and comb mode number $\eta$, see the Supplemental Material (SM). The number of ODEs equals the number of modes comprising the frequency comb; each equation follows the temporal evolution of the complex amplitude (magnitude and phase) of a single mode. In the ODEs picture, each optical comb tooth can be thought of as a nonlinear oscillator coupled to other oscillators (comb teeth). In Eq.~(\ref{eq:LLE}), $\psi(\theta, \tau)$ is the normalized field envelope, $\tau = t\Delta\omega_0/2$ is the normalized time with $\Delta\omega_0$ the resonance linewidth for the cavity mode closest to the pump (the pumped resonance) and $t$ the laboratory time, $\alpha = -2(\omega_{\mathrm{P}}-\omega_0)/\Delta\omega_0$ is the normalized detuning between the pump laser frequency $\omega_{\mathrm{P}}$ and the cold-cavity pumped resonance frequency $\omega_0$, $d_2 = -2D_2/\Delta\omega_0$ is the normalized second-order dispersion parameter, $D_2$ being the cavity second-order dispersion coefficient, and $F$ is the normalized pump amplitude. The field envelop $\psi$ and the pump amplitude $F$ are normalized to the sideband generation threshold such that the comb generation threshold in Eq.~(\ref{eq:LLE}) is equal to unity \cite{chembo2010modal}. In Eq.~(\ref{eq:CNODE}), $\tilde{a}_\eta=a_\eta(\tau) \exp[\imi\phi_\eta(\tau)]$ is the complex-valued comb tooth amplitude for mode $\eta$ with magnitude $a_\eta(\tau)$ and phase $\phi_\eta(\tau)$, $\tilde{F}_\eta(\tau)$ is the Fourier transform of $F$ and equals $\delta_{0\eta}F_{\mathrm{P}}\exp(\imi\phi_{\mathrm{P}})$ for CW pumping, $\delta_{pq}$ (for integers $p$ and $q$) is the Kronecker delta, and $\eta_{lmn}=l-m+n$. All mode numbers $\eta$ are define relative to the pumped mode. For a soliton, $\mathord{\eta\in\{0, \pm 1, \dots, \pm N\}}$ while for Turing rolls $\mathord{\eta\in\{0, \pm\mu, \pm2\mu, \dots, \pm N\mu\}}$, where the integer $\mu \geq 1$ is the mode number at which MI gain peaks.\newline
\indent When driven by a CW pump, experiments and numerical simulations suggest that for stable solutions, the magnitude of the pumped mode is much larger than that of the other modes and that in the absence of third- and higher-order dispersion, the magnitude spectrum of these solutions are symmetric with respect to the pumped mode $\eta=0$ \cite{saha2013modelocking,herr2014temporal} (see, e.g., the inset curves $a_\eta^2$ vs. mode number in Fig.~\ref{fig:prelim:phasealigned}). Therefore, we exploit the symmetry of the magnitude spectrum, adopt a perturbative approach (with $a_\eta$ for $\eta\ne0$ as the small perturbation parameters), and following Ref.~\cite{wen2014self} simplify Eq.~(\ref{eq:CNODE}) by keeping only terms with at least one contribution from the pumped mode $a_0$ in the triple summations \cite{taheri2015anatomy}. The magnitude and phase equations for the pumped mode include no linear contributions from $a_{\eta\ne0}$ (corrections are proportional to $a_\eta^2$, $\eta\ne0$), and their solutions settle on a fast time scale to the equilibrium intra-cavity field $\psi_\mathrm{e}=a_0\exp(\imi\phi_0)$; subsequently, $a_0$ and $\phi_0$ can be treated as constants (\cite{taheri2015anatomy}, also see SM).\newline
\indent Equations of motion for the magnitudes $a_\eta(\tau)$ and phases $\phi_\eta(\tau)$ are readily found from Eq.~(\ref{eq:CNODE}). The equation for the temporal evolution of the centered phase averages $\zeta_\eta=\bar{\phi}_\eta-\phi_0$, where the phase average $\bar{\phi}_\eta=(\phi_\eta+\phi_{-\eta})/2$ is centered to the pumped mode phase $\phi_0$, can be found using the equations for $\phi_{\pm\eta}$ and $\phi_0$. To lowest non-zero order in $a_{\eta\ne 0}$, this equation can be integrated directly to give
\begin{equation}\label{eq:antisym}
\tan{\zeta_\eta}=\sqrt{\abs*{\frac{C+2}{C}}}\tanh[\sqrt{|C(C+2)|}a_0^2(\tau-\tau_0)].
\end{equation}
Here $C=d_2\eta^2/2a_0^2-F_\mathrm{P}\sin(\phi_\mathrm{P}-\phi_0)/a_0^3$, $\phi_\mathrm{P}$ and $F_\mathrm{P}$ are the phase and normalized magnitude of the pump, and $\tau_0$ accounts for constants of integration (or initial conditions). Equation~(\ref{eq:antisym}) holds when $|2a_0^2-\alpha+d_2\eta^2/2|<a_0^2$, a condition that is automatically satisfied when MI gain exists (see SM).  Because the hyperbolic function approaches unity as $\mathord{\tau\to\infty}$, $\bar{\phi}_\eta$ reaches the same constant irrespective of the initial conditions.  Since $\bar{\phi}_\eta$ is fixed, each pair of phases $\phi_{\pm \eta}$ must take values symmetrically located relative to the same value. We will refer to this as ``\emph{anti-symmetrization}'' of the phases, following the terminology used in \cite{wen2014self}.  Once established, phase anti-symmetrization means each centered phase average $\zeta_\eta$ can be treated as a constant to first order in $a_{\eta\ne0}$. \newline
\indent The equations of motion for the phase differences $\Delta_\eta=(\phi_\eta-\phi_{-\eta})/2$, 
\begin{equation}\label{eq:pl}
\dot{\Delta}_\eta=\frac{a_0}{a_\eta}\sum\nolimits_l K(l,\eta) \sin(\Delta_l+\Delta_{\eta-l}-\Delta_\eta),
\end{equation}
are found by combining the equations for each $\pm\eta$ mode pair (see SM). In Eq.~(\ref{eq:pl}), the over-dot indicates time derivative, and $K(l,\eta)=a_l a_{\eta-l} \{2 \sin(\zeta_\eta-\zeta_{\eta-l}+\zeta_l)+\sin(\zeta_\eta-\zeta_{\eta-l}-\zeta_l)\}$ is the coupling coefficient for the pump--non-degenerate interaction of comb teeth $\tilde{a}_0$, $\tilde{a}_l$, $\tilde{a}_{\eta-l}$, and $\tilde{a}_\eta$. A family of fixed point solutions of Eq.~(\ref{eq:pl}) is $\Delta_\eta=s_0\eta+k\pi$, where $s_0$ is a constant and $k$ an integer.  These solutions imply that the phases have aligned:  the slope of the line passing through the phases of any pair of comb teeth $\eta$ and $-\eta$ will be the same and equal to $s_0$, that is $(\phi_\eta-\phi_{-\eta})/2\eta=s_0$. Figure~\ref{fig:prelim:phasealigned} shows two examples, in solitons and Turing rolls, where Eq.~(\ref{eq:LLE}) has been integrated numerically using the split-step Fourier transform method for a typical microresonator.

\begin{figure}[t]
\centering
\includegraphics[width=0.48\textwidth]{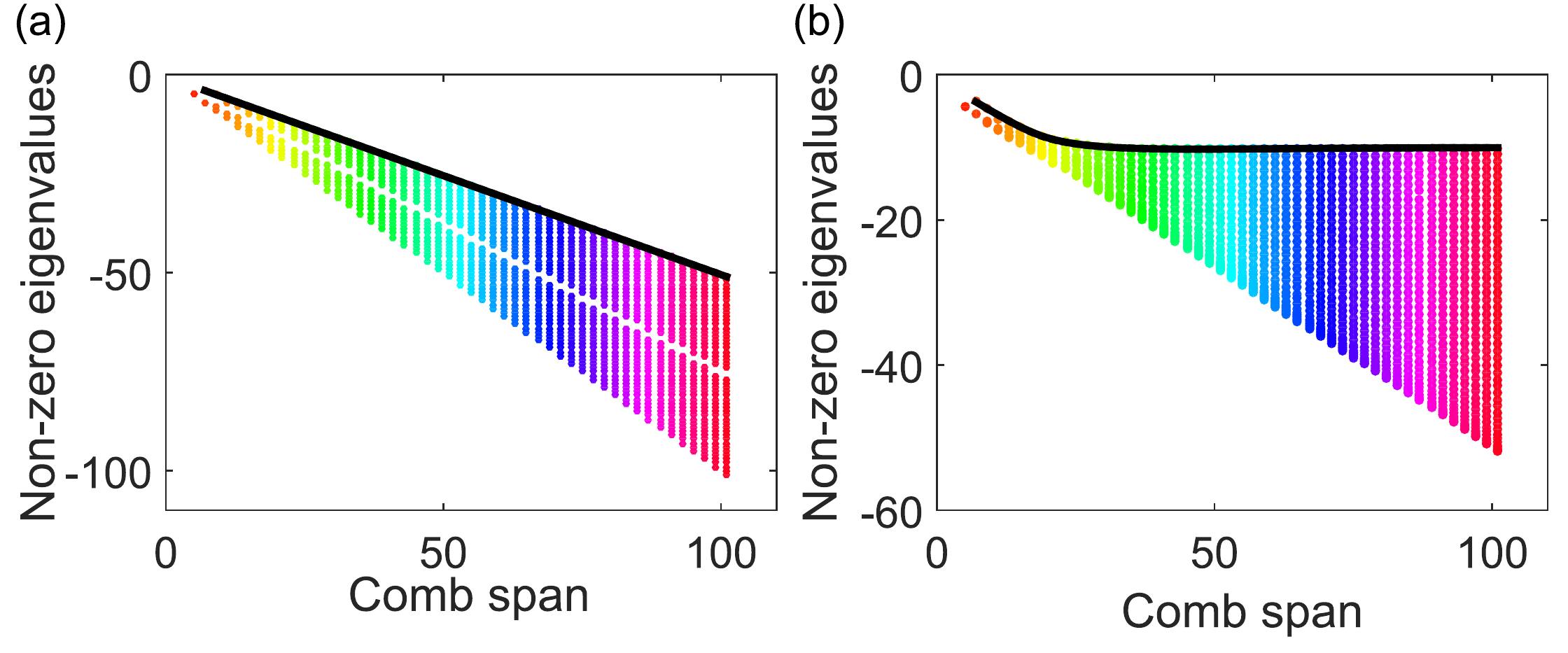}
\caption{\label{fig:prelim:eigs}Non-zero eigenvalues of the equilibrium (the Jacobian matrix $\bm{\mathrm{J}}$) versus comb span for Eq.~(\ref{eq:pl}) for (a) uniform and (b) mode-number--dependent comb teeth magnitude profile of $\mathord{a_\eta\propto\exp(-k_0|\eta|)}$, ($k_0=0.1$). The negative eigenvalue of smallest magnitude (black curve) increases in size with increasing comb span for constant magnitudes, but reaches a constant for the realistic comb magnitude profile.}
\end{figure}

Next, we consider the stability of these states.  For simplicity, we take $k=0$.  (Stability analysis for $k\ne0$ follows in a similar way.)  The linear stability of a frequency comb with $2N+1$ phase-locked teeth, with mode numbers $\mathord{\eta\in\{0,\pm 1,\pm 2,…,\pm N\}}$, is found using Eq.~(\ref{eq:pl}). We temporarily ignore the dependence of the comb teeth magnitudes on the mode number and take $a_\eta=a$. (The effect of the mode number dependence of the comb teeth magnitudes will be included shortly.) After phase locking, the centered phase averages $\zeta_\eta$ reach a steady-state value which is independent of mode number $\eta$ since the phases $\phi_\eta$ lie on a straight line. Therefore, the coupling coefficients in Eq.~(\ref{eq:pl}) are all equal, i.e., $K(l,\eta)=K > 0$. The Jacobian matrix $\bm{\mathrm{J}}$ and its eigenvalues can be expressed in closed form for any $N$ (see SM). Except for one zero eigenvalue forced by the rotational symmetry of the LLE, all of the eigenvalues are negative and real, indicating asymptotic stability of the synchronized state. Figure~\ref{fig:prelim:eigs}(a) shows the non-zero eigenvalues of the equilibrium for increasing comb span ($2N+1$). It is seen that the eigenvalue closest to zero grows more negative with increasing comb span. Hence, for the constant comb amplitude case, a wider comb demonstrates superior stability.

To investigate the effect of a non-constant comb amplitude profile, we set $\mathord{a_\eta\propto\exp(-k_0|\eta|)}$. This profile assumes a linear decay (in logarithmic scale) of the comb teeth magnitude with slope $-20k_0$ dB per increasing mode number by unity, (see, e.g., the insets $a_\eta^2$ vs. mode number in Fig.~\ref{fig:prelim:phasealigned}).  Though not analytically tractable, we find numerically that the eigenvalues of $\bm{\mathrm{J}}$ all have negative real part (except for the single zero eigenvalue forced by symmetry). Figure~\ref{fig:prelim:eigs}(b) shows the eigenvalue spectrum vs. increasing combs span for $\mathord{a_\eta\propto\exp(-k_0|\eta|)}$. Note that as the comb span increases, the smallest magnitude eigenvalue becomes bounded and almost independent of $N$ (black curve in Fig.~\ref{fig:prelim:eigs}(b)). Therefore, the stability of the comb \emph{does not improve}---nor does it degrade---with increasing comb span when the mode number dependence of the comb teeth magnitude is taken into account. Pfeifle \emph{et al}. \cite{pfeifle2015optimally} showed that in the presence of pump magnitude and frequency noise, solitons are less robust than Turing rolls in the same microresonator with comparable pump power. Our results suggest that the superior stability of Turing rolls does not originate from their smaller number of comb teeth compared to solitons. Rather, it is linked to the presence of MI gain, which is responsible for the generation of Turing rolls from vacuum fluctuations. We note that Eq.~(\ref{eq:pl}) does not explicitly include the effect of MI gain; this influence is reflected through the coupling coefficients $K(l,\eta)$.
\begin{figure*}[htbp]
\centering
\includegraphics[width=0.7\textwidth]{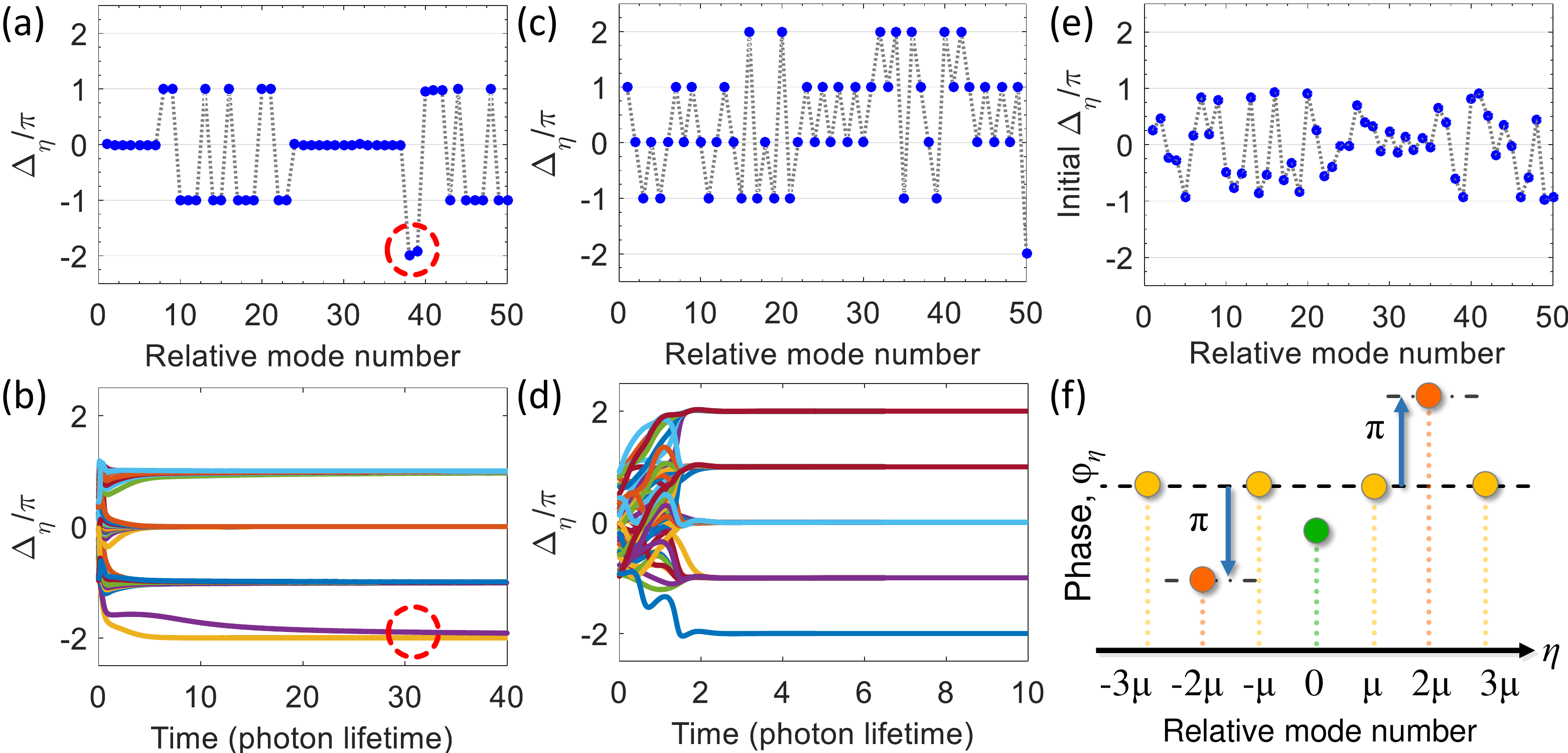}
\caption[Numerical solutions of Eq.~(\ref{eq:pl}) for uniform as well as mode-number--dependent comb teeth magnitude spectra, and the emergence of phase steps.]{\label{fig:prelim:numsol}Numerical solutions of Eq.~(\ref{eq:pl}) for uniform and mode-number--dependent comb teeth magnitude spectra, and the emergence of $\pi$ phase steps. (a) Sample steady-state solution of Eq.~(\ref{eq:pl}) for a comb with 101 teeth, with uniform magnitudes and random initial conditions for the phase differences (shown in (e)). (b) The temporal evolution of the phase differences shown in (a). Most of the phase differences settle to integer multiples of $\pi$, but some deviations may arise (dotted red circles). (c,d) Same as (a,b) but for comb magnitude profile of $\mathord{a_\eta\propto\exp(-k_0|\eta|)}$, ($k_0=0.1$). All of the phase differences $\Delta_\eta$ settle to integer multiples of $\pi$. In (a,c), only the $\pi$ phase steps are physically significant. The $2\pi$ steps have not been removed (e.g., through unwrapping the phases) to better illustrate the correspondence of (a,c) with (b,d). (e) The phase differences at the onset of integration (initial conditions) for (a-d). (f) Schematic illustrating $\pi$ steps in the phase profile $\phi_\eta$ of a comb resulting from steps in the phase differences $\Delta_\eta$. Comb teeth symmetrically positioned around the pumped mode ($\eta$ and $-\eta$) will show $\pi$ phase steps, with one phase increasing as its counterpart decreases. Such phase steps do not change the phase averages $\bar{\phi}_\eta$.}
\end{figure*}

\begin{figure}[htbp]
\centering
\includegraphics[width=0.49\textwidth]{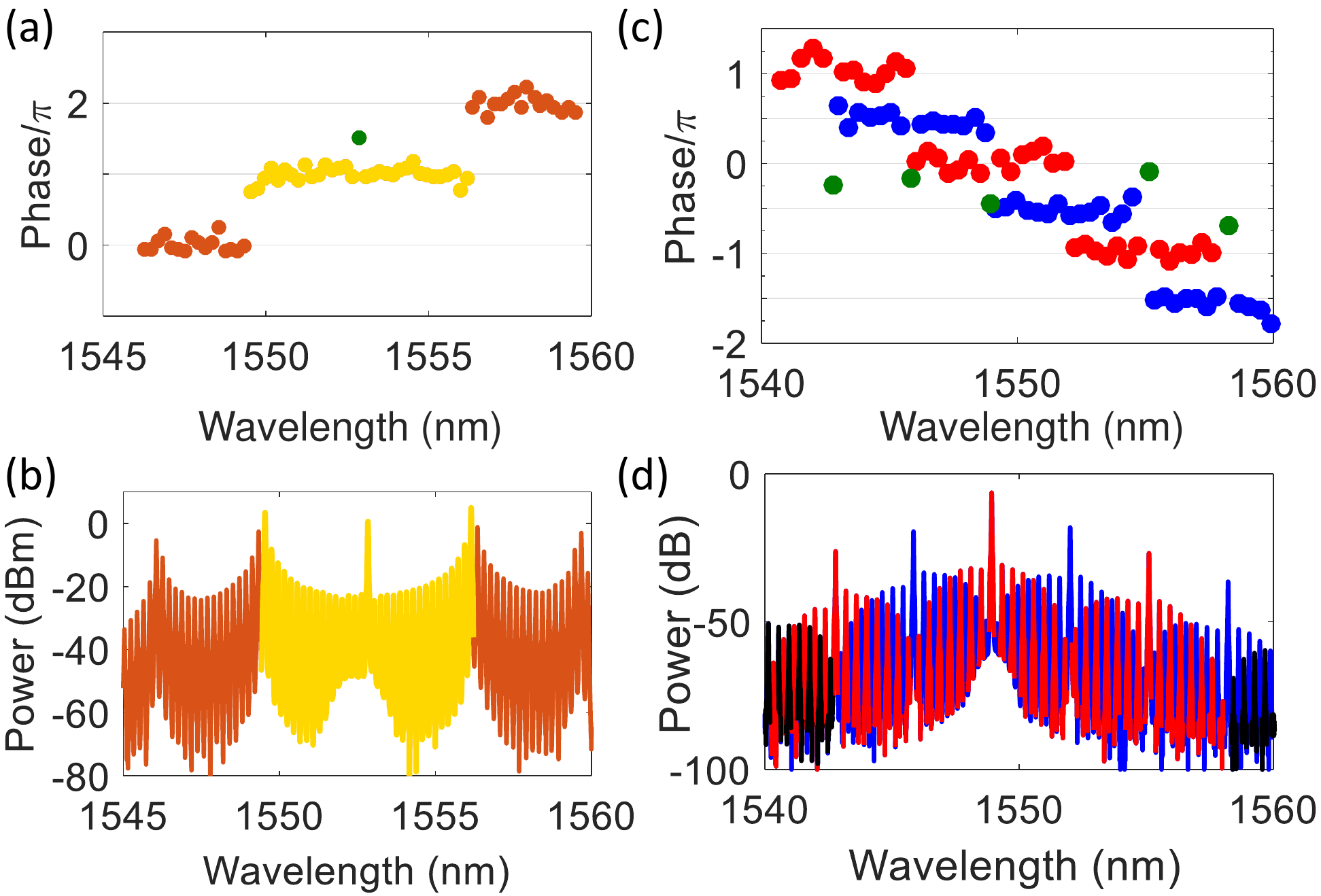}
\caption[Experimental data showing steps in the measured phase spectra of optical frequency combs.]{\label{fig:prelim:steps}Experimental data showing steps in the measured phase spectra of optical frequency combs. (a,c) are the phase spectra while (b,d) depict the power profiles of the combs. (a,b) $\pi$-steps in the phase spectrum of a stable comb; The green dot corresponds to the pumped mode phase. (c,d) $\pi$/2-steps in the phase spectrum of a stable comb; This comb is recognized as two interleaved combs (red and blue) with phases offset by $\pi/2$, each exhibiting $\pi$-phase steps as well (cf. Fig.~\ref{fig:prelim:numsol}(f)). The $\pi/2$-phase offset leads to the decoupling of the two combs, indicated by vanishing coupling coefficients, i.e., $K(l, \eta)=0$ in Eq.~(\ref{eq:pl}). The green dots correspond to the phases of the stronger comb teeth. (These plots are reproduced using data originally presented in~\cite{del2015phase}.)}
\end{figure}

We consider next numerical solutions of Eq.~(\ref{eq:pl}). Our numerous runs of numerical integration for different comb spans ($N$ from 5 to 1000) and random initial phase differences taken from a uniform distribution over the range $(-\pi, \pi]$ and for uniform comb magnitude spectrum, typically lead to $\Delta_\eta=k\pi$, ($k$ an integer). While a steady-state is always obtained, other steady-state solutions are also possible. A mode-number-dependent comb magnitude spectrum, in contrast, always leads to phase differences equal to integer multiples of $\pi$, even for those cases in which steady-state phase differences for uniform comb magnitude spectra are not equal to integer multiples of $\pi$. The reason is that a non-uniform comb magnitude profile places more strict constraints on the steady-state solution of Eq.~(\ref{eq:pl}). In Figs.~\ref{fig:prelim:numsol}(a-d), we show sample solutions found by numerically integrating Eq.~(\ref{eq:pl}) for $N=50$ phase differences (a comb with $2N+1=101$ teeth) for constant comb teeth magnitudes (Fig.~\ref{fig:prelim:numsol}(a,b)) and the non-constant magnitude profile of $\mathord{a_\eta\propto\exp(-k_0|\eta|)}$ (Fig.~\ref{fig:prelim:numsol}(c,d)). We show the steady-state solutions $\Delta_\eta$ at the end of the simulation time vs. mode number as well as the the evolution of the phase differences with time. The initial values of $\Delta_\eta$, $\mathord{\eta\in\{1,2,…,50\}}$, in both cases is shown in Fig.~\ref{fig:prelim:numsol}(e). While most of the steady-state phase differences in Fig.~\ref{fig:prelim:numsol}(a) are integer multiples of $\pi$, some of them deviate from these values (the dotted red circles). For the non-uniform magnitude spectrum, however, steady-state phase differences are all integer multiples of $\pi$, as seen in Fig.~\ref{fig:prelim:numsol}(c). \newline
\indent The $\pi$ phase steps in the phase differences $\Delta_\eta$ imply similar steps in the phases $\phi_\eta$. To show this, we assume a set of solutions $\Delta_\eta=s_0\eta$ is known and try to find another set based on it. It may seem that any constant $x$ can be added to the phases of comb teeth symmetrically positioned with respect to the pumped mode (i.e., $\phi_{\pm\eta}\to\phi_{\pm\eta}+x$) without affecting the solution. Unfortunately, this alters the phase averages $\bar{\phi}_\eta$ and so invalidates the stability analysis presented earlier. However, we can generate new stable phase-locked solutions by considering anti-symmetric changes of the phases, i.e., $\phi_\eta\to\phi_\eta\pm x$ and $\phi_{-\eta}\to\phi_{-\eta}\mp x$, which means $2\Delta_\eta=\phi_\eta-\phi_{-\eta}\pm 2x=2s_0\eta +2k\pi$, and hence $x=\pm \pi$ (recall that $k$ is an integer). This demonstrates that the appearance of $\pi$ steps in the phase spectrum of stable phase-locked frequency combs is permissible, as shown schematically in Fig.~\ref{fig:prelim:numsol}(f). Indeed, such phase steps have been observed experimentally \cite{del2015phase} (see, e.g., Fig.~\ref{fig:prelim:steps}(a)) but have, to the best of our knowledge, remained unexplained until now. \newline
\indent Besides $\pi$ phase steps, $\pi/2$ phase steps \cite{del2015phase} have also been observed in experiments. These phase steps also can be explained within the framework of our model, as follows. A comb with $\pi/2$ phase steps is in fact two interleaved \emph{non-interacting} combs, each of which has $\pi$ steps in its phase spectrum \cite{del2015phase}. These two combs do not interact as a result of the $\pi/2$ offset between their phase spectra because this phase offset causes the coupling coefficients between their comb teeth, $K(l, \eta)$, to vanish. To clarify this point, we refer to Fig.~\ref{fig:prelim:steps}(c,d) where comb teeth labeled with $\mathord{\eta_\mathrm{A}\in\{0,2,4,6,…,14\}}$ (red) share a constant phase, while those with mode numbers $\mathord{\eta_\mathrm{B}\in\{1,3,5,…,13\}}$ (blue) share another phase, different from that of the former group by $\pi/2$, such that $\zeta_1=\bar{\phi}_1-\phi_0=\pi/2$ and $\zeta_{\eta_\mathrm{A}}-\zeta_{\eta_\mathrm{B}}=\pi/2$ (recall that the value of $\zeta_{\eta}$ is independent of the mode number for a stable comb, cf. Fig.~\ref{fig:prelim:numsol}(f)). As a result, the coupling coefficient $K(\eta, \eta+1)$ is zero because $\zeta_{\eta+1}\pm\zeta_{\eta}=\pm\pi/2$, and so there is no coupling between modes $\eta$ and $\eta+1$ of the comb for any $\eta$ (see Eq.~(\ref{eq:pl})). It is worth noting that the frequency combs with phase steps in \cite{del2015phase} were obtained through tuning the laser pump into resonance, which alters the MI gain profile and sweeps its peak.\newline
It has been argued that passing through the chaotic state is necessary for microcomb soliton formation \cite{lamont2013route}. The foregoing analysis suggests a way of understanding this: passage through the chaotic state serves to provide the system with a large pool of initial conditions, which increases the odds of getting peaks that will then grow into solitons. Numerical simulations of Eq.~(\ref{eq:pl}) suggest that with increasing comb span and non-uniform comb magnitude spectrum, chances of getting groups of phase-locked comb teeth, or weak pulses, will increase. These weak pulses will then grow into the modes of the nonlinear system (i.e., the solitons). \newline
\indent Finally, although we have compared our theoretical results with microresonator-based frequency comb experiments, they should also apply to mode-locked laser systems.  In 2002, Gordon and Fisher developed a many-body statistical mechanical theory to describe the onset of laser pulsations as a first order phase transition, treating the modes as the elementary degrees of freedom \cite{gordon2002phase}.  Their ordered collective state is analogous to our synchronized dynamical attractor.  Now, Eq.~(\ref{eq:pl}) roots in the cubic nonlinear term in the LLE, and the same nonlinearity appears in the master equation for passive mode locking based on a saturable absorber, which approximates the absorber with a cubic nonlinearity \cite{haus2000mode}. We therefore expect the same dynamical mechanism to be responsible for the creation of sharp pulses in passively mode-locked lasers, despite the different physical source of optical gain (population inversion and stimulated emission rather than parametric amplification).  What matters is the fundamental link between spatiotemporal pulse formation and mode synchronization. \newline
\indent The generic reduced nonlinear oscillator model introduced in this work clearly demonstrates the fundamental link between mode synchronization and spatiotemporal pulse formation in Kerr-nonlinear media. This model admits attracting fixed point solutions corresponding to stable cavity solitons and Turing patterns, permits analyzing their stability in a unified scheme, and explains phase jumps observed in recent phase measurements of stable optical frequency combs. It also provides insight into the role of chaos and parametric gain in the generation of solitons and Turing rolls. This insight can be utilized towards devising novel techniques for controlled formation of robust pulses in optical microresonators.
\begin{acknowledgments}
K.W. and H.T. thank Brian Kennedy for useful discussions.  K.W. also thanks Henry Wen and Steve Strogatz for generously discussing the details of their results reported in \cite{wen2014self}.  H.T. was supported by the Air Force Office of Scientific Research Grant No.~2106DKP.
\end{acknowledgments}

\bibliography{references}

\end{document}